\documentclass[prd,reprint,amsmath,amssymb,preprintnumbers,superscriptaddress,nofootinbib]{revtex4-1}

\usepackage{slashed}
\usepackage{subfigure}
\usepackage[colorlinks]{hyperref}
\usepackage[dvipdfmx]{graphicx}
\usepackage{color}

\begin{document}

\preprint{CTPU-17-04}

\title{\boldmath Sterile neutrino dark matter from right-handed neutrino oscillations}

\author{Kenji Kadota}
\affiliation{Center for Theoretical Physics of the Universe, Institute for Basic Science (IBS), Daejeon 34051, Korea}
\author{Kunio Kaneta}
\affiliation{Center for Theoretical Physics of the Universe, Institute for Basic Science (IBS), Daejeon 34051, Korea}

\begin{abstract}
  \noindent
  We study a scenario where sterile neutrino (either warm or cold) dark matter (DM) is produced through (nonresonant) oscillations among right-handed neutrinos (RHNs) and can constitute the whole DM in the Universe, in contrast to the conventional sterile neutrino production through its mixing with the left-handed neutrinos.
  The lightest RHN can be sterile neutrino DM whose mixing with left-handed neutrinos is sufficiently small while heavier RHNs can have non-negligible mixings with left-handed neutrinos to explain the neutrino masses by the seesaw mechanism. We also demonstrate that, in our scenario, the production of sterile RHN DM from the decay of a heavier RHN is subdominant compared with the RHN oscillation production due to the X-ray and small-scale structure constraints. 

\end{abstract}
\maketitle

\section{Introduction}
\label{sec:introduction}

While it has been established that neutrinos are massive due to the discovery of neutrino oscillations \cite{Fukuda:1998mi,Ahmad:2002jz}, their precise properties, however, are still under active investigation.
An analogous (and even more perplexing) story applies to dark matter (DM) whose nature remains unknown despite the ever-growing evidence for its existence from the astrophysical observables. 
An intriguing possibility regarding these mysteries would be to introduce right-handed neutrinos (RHNs), which can address not only the neutrino mass and DM but also their potential roles in the inflation and baryon asymmetry production \cite{seesaw,Asaka:2005an,Fukugita:1986hr,Asaka:2005pn,Canetti:2012kh,Ibe:2015nfa,kk2}. 

We, in this article, seek a possibility for a sterile RHN to make up the whole DM in the Universe and, in particular, propose the new production mechanism of sterile RHN DM through the mixing among RHNs. 
This is in contrast to the conventional mechanisms requiring the sterile RHN DM to couple to left-handed neutrinos which suffer from the severe tension between the bounds from the X-ray observation and the small-scale structure data \cite{Dodelson:1993je,Shi:1998km,Tremaine:1979we,Boyarsky:2005us,Horiuchi:2013noa,ir2017}. These constraints, however, heavily depend on their production mechanisms and many possibilities have been explored to produce the desired DM abundance in addition to the conventional nonresonant/resonant active-sterile neutrino conversion mechanisms \cite{Asaka:2006ek,Bezrukov:2009th,kenji,Anisimov:2008gg,Adhikari:2016bei,Asaka:2005pn,Canetti:2012kh,Ibe:2015nfa,kk2}.

Our scenario is distinguishable from such alternative scenarios in that it still uses a simple oscillation between the thermal heavy RHN and DM, and yet it demonstrates the totally different features from the Dodelson-Widrow scenario such as the occurrence of the production peak above/around the electroweak which is of great advantage in circumventing the Lyman-$\alpha$ bounds due to the redshifting of DM momentum.   
After outlining our setup in Sec. II, we illustrate our scenario in Sec. III for a simple example of two RHNs. Section IV then demonstrates the concrete realization where we introduce a RHN mass matrix whose off-diagonal term can arise from the scalar field vacuum expectation value so that we can explain the light neutrino masses by the seesaw mechanism while avoiding the tight X-ray bounds. Section V is devoted to the discussion/conclusion. 

\section{Setup}
\label{sec:setup}
The Lagrangian we study is the standard model (SM) with three Majorana RHNs, given by ${\cal L} = {\cal L}_{\rm SM} + {\cal L}_N$ where ${\cal L}_{\rm SM}$ is the SM Lagrangian and ${\cal L}_N$ reads
\begin{eqnarray}
  	\overline \nu_R i\slashed{\partial} \nu_R - 
	\left[
		\nu_R^{c}{}^T y_\nu L H - \frac{1}{2}\nu_R^{c}{}^T{\cal M}_N\nu_R^c + {\rm H.c.}
        	\right], \label{eq:lagrangian}
\end{eqnarray}
where $H, L$, and $\nu_R$ are, respectively, the Higgs doublet, lepton doublet and RHN. For simplicity, we concentrate on the case of three RHNs.

We begin with the field basis where $y_\nu y_\nu^\dagger$ is diagonal, denoted as $y_\nu^{\rm diag}$ so that $y_\nu^{\rm diag} y_\nu^{{\rm diag}\dagger}$ becomes a $3\times 3$ diagonal matrix. ${\cal M}_N$
 is, in general, a nondiagonal matrix, which we call the interaction basis.
A familiar seesaw mechanism for the mass of left-handed neutrino $\nu_L$ reads, in terms of its Dirac mass $m_D^{\rm diag} = y_\nu^{\rm diag} v$ with $v=\langle H\rangle$, ${\cal M}_\nu = m_D^{\rm diag} {}^T {\cal M}_N^{-1} m_D^{\rm diag}$ which can be diagonalized as ${\cal M}_\nu^{\rm diag} = U_L^T {\cal M}_\nu U_L$ ($U_L$ is the Pontecorvo-Maki-Nakagawa-Sakata matrix.\footnote{Throughout this article, we take the charged lepton Yukawa coupling to be diagonal.}).
The neutrino mass eigenstates are
\begin{eqnarray}
	\left[
		\begin{array}{c}
			\nu_L\\
			\nu_R^c
		\end{array}
	\right]
	&=&
	U
	\left[
		\begin{array}{c}
			\nu\\
			N^c
		\end{array}
	\right],
	\quad
	U \simeq
	\left[
		\begin{array}{cc}
			1 & \theta^\dagger \\
			-\theta & 1
		\end{array}
	\right]
	\left[
		\begin{array}{cc}
			U_L & \\
			 & U_R^*
		\end{array}
	\right],
	\label{eq:U}
\end{eqnarray}
where $\theta\equiv {\cal M}_N^{-1}m_D^{\rm diag}$ and $U_R$ is a unitary matrix defined to diagonalize ${\cal M}_N$ as ${\cal M}_N^{\rm diag}=U_R^\dagger {\cal M}_N U_R^*$.
By taking the rotation of Eq.~(\ref{eq:U}), the Yukawa coupling $y_\nu$ is in general a nondiagonal matrix while the neutrino masses, ${\cal M}_\nu$ and ${\cal M}_N$, are simultaneously diagonalized. We call this field basis the mass basis.
Thus, we obtain
\begin{eqnarray}
	y_\nu^{\rm diag}y_\nu^{\rm diag}{}^\dagger
	&=&
	v^{-2}
	\left[
		U_R({\cal M}_N^{\rm diag})^{1/2}R({\cal M}_\nu^{\rm diag})^{1/2}
	\right]\nonumber\\
	&&\times
	\left[
		U_R({\cal M}_N^{\rm diag})^{1/2}R({\cal M}_\nu^{\rm diag})^{1/2}
	\right]^\dagger,
	\label{eq:yuk}
\end{eqnarray}
where $R$ is an arbitrary $3\times 3$ complex orthogonal matrix satisfying $R^TR=1$~\cite{Casas:2001sr}.
The mixing between $\nu_L$ and $N$ is then parametrized by $\Theta=\theta^\dagger U_R^*$, and
\begin{eqnarray}
	\Theta^2 &\equiv&
	\Theta^\dagger\Theta = ({\cal M}_N^{\rm diag})^{-1/2}R{\cal M}_\nu^{\rm diag}R^\dagger({\cal M}_N^{\rm diag})^{-1/2}.
	\label{eq:active-sterile mixing}
\end{eqnarray}

The oscillations among RHNs can take place when their mass and interaction bases differ.
We, in the following discussions, consider three RHNs with their masses ${\cal M}_N^{\rm diag}={\rm diag}\{M_1,M_2,M_3\}$ and take $N_1$ as the lightest one so that it can play the role of DM. For the active neutrino masses, we parametrize ${\cal M}_\nu^{\rm diag} = {\rm diag}\{m_1,m_2,m_3\}$ for the normal hierarchy (NH), where $\Delta m_{21}^2 \equiv m_2^2-m_1^2 = (7.50^{+0.19}_{-0.17})\times 10^{-5}~{\rm eV}^2,
\Delta m_{31}^2 \equiv m_3^2-m_1^2 = (2.457^{+0.047}_{-0.047})\times 10^{-3}~{\rm eV}^2$
\cite{Gonzalez-Garcia:2015qrr}.
For the inverted hierarchy (IH), we take ${\cal M}_\nu^{\rm diag} = {\rm diag}\{m_3,m_1,m_2\}$ and $\Delta m_{32}^2 \equiv m_3^2-m_2^2 = (-2.449^{+0.048}_{-0.047})\times 10^{-3}~{\rm eV}^2$.
The lightest neutrino mass ($m_1$ for the NH case, and $m_3$ for the IH case) is taken as a free parameter.
In our discussions below, whenever it is not necessary to distinguish the mass orderings, $m_1$ refers to the lightest mass for brevity.

\section{DM production through RHN oscillation}
We now check if enough abundance of RHN DM $\nu_{R1}$ can be produced from the RHN oscillations. In our scenario, the heavy RHNs $\nu_{R2}$ and $\nu_{R3}$ explain the left-handed neutrino masses by the seesaw mechanism and they can have sizable neutrino Yukawa couplings to be in the thermal equilibrium at a sufficiently high temperature. $\nu_{R1}$, on the other hand, has a sufficiently small coupling to the SM species, so that its production is dominated by the conversion from heavier RHNs. For clarity of the following quantitative discussion, we focus on the $\nu_{R1}$ abundance produced only from its mixing with  $\nu_{R2}$ because $\nu_{R3}$ plays the same role as $\nu_{R2}$ in producing $\nu_{R1}$.

The relevant reactions for the $\nu_{R2}$ thermalization are the scatterings caused by Yukawa interaction, $\nu_{R2} L \leftrightarrow t Q_3,~\nu_{R2} t \leftrightarrow L Q_3,~\nu_{R2} Q_3 \leftrightarrow L t$, those involving the gauge bosons, $\nu_{R2} V \leftrightarrow H L,~\nu_{R2} L \leftrightarrow H V,~\nu_{R2} H \leftrightarrow L V$ and the decay and inverse decay $\nu_{R2}\leftrightarrow LH$ [$Q_3 (t)$ is the left (right) handed top quark, and $V$ represents the $SU(2)_L$ and $U(1)_Y$ gauge bosons].

The Boltzmann equation for $\nu_{R1}$ \cite{Dolgov:2000ew} reads
\begin{eqnarray}
	\frac{dn_{\nu_{R1}}}{dt} + 3Hn_{\nu_{R1}} = C_{\nu_{R1}}
\end{eqnarray}
where $C_{\nu_{R1}}$ represents the collision term integrated over the $\nu_{R1}$ momentum given by
\begin{eqnarray}
    C_{\nu_{R1}} &\simeq& {\cal P}(\nu_{R2}\to\nu_{R1})(\gamma_{\nu_{R2}}^{\rm col}+\gamma_{\nu_{R2}}^{\rm ID}),\\
	\gamma_{\nu_{R2}}^{\rm col} &=& \frac{T}{64\pi^4}\int^\infty_{s_{\rm min}} ds\hat\sigma\sqrt{s}K_1(\sqrt{s}/T),\\
	\gamma_{\nu_{R2}}^{\rm ID} &=& \frac{M_2^2T}{\pi^2}\Gamma(\nu_{R2}\to LH)K_1(M_2/T).
\end{eqnarray}
Here ${\cal P}$ is the oscillation probability given by ${\cal P}(\nu_{R2}\to\nu_{R1}) = \frac{1}{2}\sin^22\theta_N$ ($\theta_N$ is the mixing angle between $\nu_{R1}$ and $\nu_{R2}$), $\Gamma(\nu_{R2}\to LH)\simeq (y_\nu y_\nu^\dagger)_{22}M_2/(8\pi)$ is the decay width, and $\hat\sigma$ is the reduced cross section for the $\nu_{R2}$ collisions with the kinematical cut $s_{\rm min}$ of the Mandelstam variable $s$, and $K_1$ is the modified Bessel function of the first kind. 
\footnote{A factor 1/2 in ${\cal P}$ comes from averaging out the RHN oscillation because the oscillation timescale is much shorter than the collision timescale involving $\nu_{R2}$. More quantitatively, this averaging is justified for $T\lesssim 10^6$ GeV and/or $\Delta M^2\equiv M_2^2-M_1^2 \gtrsim 1~{\rm GeV}^2$ because $t_{\rm osc}/t_{\rm col} \sim({y_\nu^2}/{10^{-14}})({g^2}/{10^{-2}})({\rm GeV}^2/{\Delta M^2})({T}/{10^6~{\rm GeV}})^2$ where $g$ represents a gauge coupling for a relevant gauge interaction. As we will discuss later, $y_\nu^2$ of order $10^{-14}$ is required for GeV-scale RHN to reach the thermal equilibrium and it is automatically realized by enforcing the seesaw mechanism. The finite temperature effects on the RHN mixing angle $\theta_N$ are suppressed by the neutrino Yukawa couplings in our scenario and we simply consider a constant $\theta_N$ in our estimation. The cases when these approximations are not applicable are left for the future work.}

\begin{center}
\begin{figure}[t]
\includegraphics[width=0.45\textwidth]{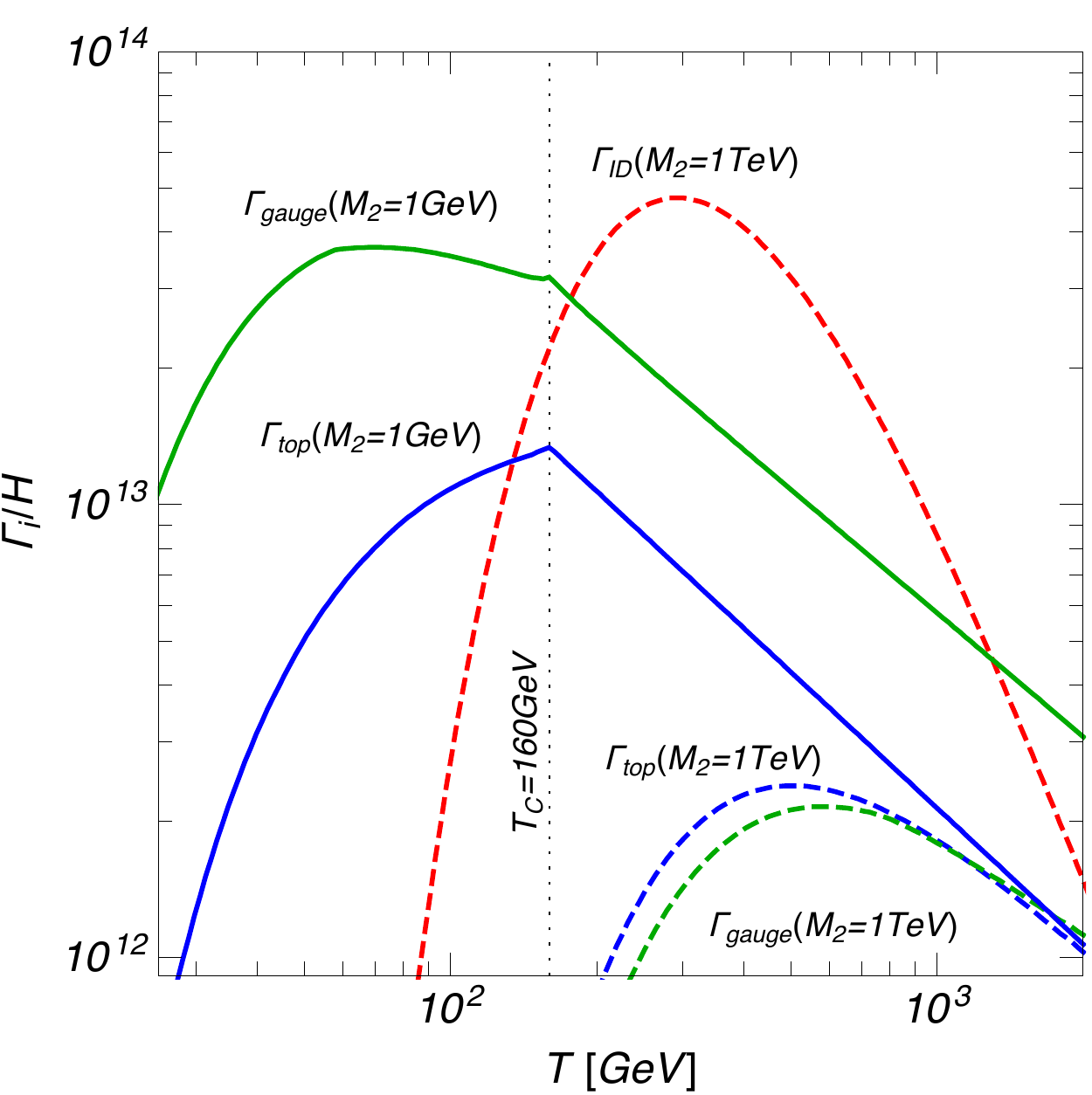}
\caption{
The ratios between the rescaled (i.e., divided by the Yukawa couplings) reaction rates and the Hubble parameter are shown (the actual reaction rates are obtained by multiplying the Yukawa couplings). The solid curves are for $M_2=1$ GeV and the dashed curves are for $M_2=1$ TeV. 
}
\label{fig:rate}
\end{figure}
\end{center}

$\nu_{R1}$ is efficiently produced when the collision terms are large.\footnote{Some of collision terms, such as $\nu_R H\to LV$, possess the infrared divergences, which are regulated by the thermal mass of the propagator in our analysis for $T>T_C$ ($T_C$ is the critical temperature of the electroweak phase transition and we take $T_C = 160$ GeV) \cite{Pilaftsis:2003gt,Besak:2012qm,DOnofrio:2014rug,DOnofrio:2015gop}.}
Figure \ref{fig:rate} shows $\Gamma_i/H$ where $\Gamma_i$ represents the rescaled reaction rates for the process $i$ by taking the neutrino Yukawa coupling as unity (so that the curves can be easily scaled by multiplying the Yukawa coupling of interest).
For illustration purpose, we define the reaction rates $\Gamma_i=\gamma^{\rm col}_{\nu_{R2}}(i)/n_\gamma$, where $n_\gamma=2T^3/\pi^2$ is the radiation number density and $\gamma^{\rm col}_{\nu_{R2}}(i)$ are the collision terms involving the gauge bosons [$\gamma^{\rm col}_{\nu_{R2}}({\rm gauge})$] and the top quarks [$\gamma^{\rm col}_{\nu_{R2}}({\rm top})$]. The inverse decay rate is given by $\Gamma_{\rm ID}=\gamma^{\rm ID}_{\nu_{R2}}/n_\gamma$.
The figure shows the plots for $M_2=1$ GeV (solid) and for $M_2=1$ TeV (dashed), and we note that the inverse decay takes place only for the latter because of the kinematics, namely, the (inverse) decay is available only for $M_2\gtrsim M_h$ with $M_h$ being the Higgs mass. The actual reaction rates can be obtained by multiplying these rescaled reaction rates by $(y_\nu y_\nu^\dagger)_{22}$. We can see, from Fig. \ref{fig:rate}, that $N_2$ can reach the thermal equilibrium ($\Gamma_i/H \gtrsim 1$) when $(y_\nu y_\nu^\dagger)_{22}$ is larger than ${\cal O}(10^{-13})$ for $M_2=1 - 10^3$ GeV, which is also in the desired numerical range to explain the neutrino masses by the seesaw mechanism.
\vspace*{-7mm}

The produced $\nu_{R1}$ (interaction state) constitutes the DM $N_1$ (mass eigenstate), \footnote{The produced $\nu_{R1}$ is composed of $N_1$ and $N_2$ which propagate with different velocities. As the $\nu_{R1}$ energy gets redshifted, these two mass states are eventually well separated and thus $\nu_{R1}$ is expected to mostly develop the $N_1$ component as long as $M_1\ll M_2$, although the oscillation property may call for a careful study \cite{Akhmedov:2012uu}.} and the current $N_1$ relic number density can be estimated, in terms of the yield parameter $Y_{N_1}\equiv n_{N_1}/s$ ($s$ is the entropy density), by integrating the Boltzmann equation from $T_{\rm RH}$, the reheating temperature, to the current temperature $T=T_0$
\begin{eqnarray}
	Y_{N_1}^0\equiv Y_{N_1}(T=0) = \int^\infty_0dT {\cal P}(\nu_{R2}\to\nu_{R1})\frac{\gamma_{\nu_{R2}}}{sHT},
\end{eqnarray}
where we have taken the limits $T_{\rm RH}\to\infty, T_0\to0$, and $\gamma_{\nu_{R2}}\equiv\gamma_{\nu_{R2}}^{\rm col}+\gamma_{\nu_{R2}}^{\rm ID}$. 
The corresponding DM density can then be estimated in terms of the yield parameter
\begin{eqnarray}
	\Omega_{N_1}h^2 &\simeq&
	0.12
	\left[
		\frac{\sin^22\theta_N}{8.8\times10^{-3}}
	\right]
	\left[
		\frac{|y_\nu^{\rm diag}|^2_{22}}{10^{-13}}
	\right]
	\left[
		\frac{M_1}{\rm keV}
	\right]
	\left[
		\frac{\tilde Y_{N_1}^0}{10^{12}}
	\right],\nonumber\\
\end{eqnarray}
where $\tilde Y^0_{N_1}$ is the rescaled yield parameter, defined by factoring out the oscillation probability and the Yukawa coupling, $\tilde Y^0_{N_1} \equiv Y^0_{N_1}/({\cal P}(\nu_{R2}\to\nu_{R1}) (y_\nu y_\nu^\dagger)_{22})$.
We found the following simple fitting formula to grasp the characteristic features of the DM abundance in our scenario
\begin{eqnarray}
	\log_{10} \tilde Y^0_{N_1} 
	&\simeq& 
	12.8 \quad (M_2\lesssim M_h)\nonumber\\
	&\simeq&
	13.3-(1/2)\log_{10}(M_2/M_h) \quad (M_2\gtrsim M_h).\nonumber\\
\end{eqnarray}
This behavior matches our expectation because, as emphasized in referring to Fig. \ref{fig:rate}, the most efficient production occurs when the production rate reaches maximal with respect to the Hubble expansion rate. 
$\tilde Y^0_{N_1}$ is hence little dependent on $M_2$ when $M_2$ is smaller than $M_h$, because $N_2$ is dominantly produced via the inverse decay in this case, and thus the temperature at which the production rate becomes maximal is at $T\simeq M_h$.
For $M_2\gtrsim M_h$, on the other hand, the SM particles possess the thermal mass and the production rate becomes maximal around $T\sim M_2$, which leads to some power dependence of the yield parameter on $M_2$.
This is illustrated through a concrete example in the next section.

\section{Benchmark model}
\label{bm}

\begin{center}
\begin{figure}[t]
\includegraphics[width=0.45\textwidth]{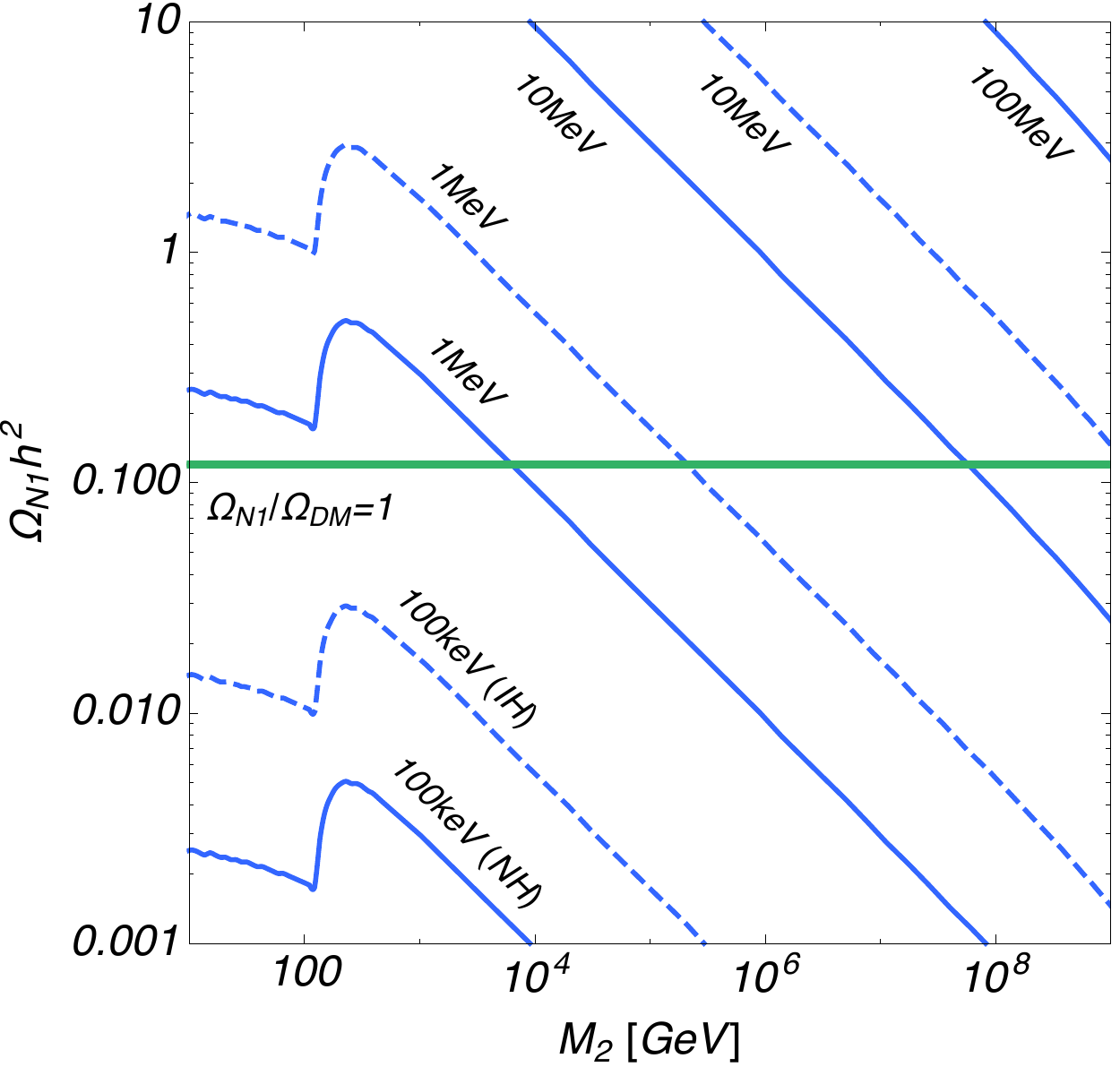}
\caption{
  The $N_1$ relic abundance is shown as a function of $M_2$ by varying $M_1$ from 100 keV to 100 MeV. The solid and dashed curves show the NH and IH cases, respectively.
}
\label{fig:BM}
\end{figure}
\end{center}
We here discuss a possible realization of our scenario.
Let us begin with a simple mass matrix given by
\begin{eqnarray}
	{\cal M}_N &=&
	\left[
		\begin{array}{ccc}
			M_0&m&\\
			m&M_2&\\
			&&M_3
		\end{array}
	\right],
\end{eqnarray}
where $m$ and $M_0$ are taken to be $M_0\lesssim m \ll M_2, M_3$.
${\cal M}_N$ is then diagonalized as ${\cal M}_N^{\rm diag}={\rm diag}\{M_1,M_2,M_3\}$ with $M_1 \simeq M_0-m^2/M_2$ by using $U_R$ which reads
\begin{eqnarray}
	U_R^* &\simeq&
	\left[
		\begin{array}{ccc}
			1&\theta_N&\\
			-\theta_N&1&\\
			&&1
		\end{array}
	        \right]
        , \theta_N=m/M_2.
	\label{eq:UR1}
\end{eqnarray}
The resultant $N_1$ abundance in the NH case is then given by
\begin{eqnarray}
	\Omega_{N_1}h^2 &\simeq&
	0.12
	\left[
		\frac{m_2}{0.01~{\rm eV}}
	\right]
	\left[
		\frac{M_1}{\rm keV}
	\right]
	\left[
		\frac{(m/5~{\rm GeV})^2}{M_2/100~{\rm GeV}}
	\right]
	\left[
		\frac{\tilde Y_{N_1}^0}{10^{13}}
	\right],\nonumber\\
\end{eqnarray}
while, in the IH case, $m_2$ should be replaced by $m_1$.
In the case of $M_0 \ll m^2/M_2$, we can take $\theta_N \simeq (M_1/M_2)^{1/2}$ due to $M_1\simeq m^2/M_2$, and thus we obtain
\begin{eqnarray}
	\Omega_{N_1}h^2 &\simeq&
	0.12
	\left[
		\frac{m_2}{0.01~{\rm eV}}
	\right]
	\left[
		\frac{M_1}{0.52~{\rm MeV}}
	\right]^2
	\left[
		\frac{\tilde Y_{N_1}^0}{10^{13}}
	\right].
	\label{eq:Oh2_NH}
\end{eqnarray}
For this simplified case, Fig. \ref{fig:BM} shows $\Omega_{N_1}h^2$ as a function of $M_2$ for various $M_1$ taken from 100 keV to 100 MeV in both the NH and IH cases which are depicted by solid and dashed curves, respectively.
\footnote{It should be noted that, in Fig.~\ref{fig:BM}, $t_{\rm osc}/t_{\rm col} \ll 1$ is achieved for $T\lesssim 10^6\times M_2$ even in the large $M_2$ region, so that a factor 1/2 in ${\cal P}$ by averaging out the RHN oscillation is justified.}
The green band in the figure indicates the observed value of the DM abundance given by $\Omega_{\rm DM}h^2=0.1197\pm0.0022$ \cite{Ade:2015xua}.

On the other hand, since $\Theta^2_{11}$ depends on $\theta_N$ and we need a relatively large $\theta_N$ for our scenario to work, the $N_1$ is subject to the X-ray constraint given by $\Theta^2_{11} \lesssim 10^{-5}({\rm keV}/M_1)^5$ \cite{Boyarsky:2005us}.
One may simply expect that the X-ray bound is easily circumvented because the Yukawa coupling of $\nu_{R1}$ can be negligibly small. We, however, point out that the light RHN can decay into the SM particles through its oscillation to a heavier RHN. 
In our current setup, we obtain $\Theta^2_{11} = M_1^{-1} (m_1|R_{11}|^2 + m_2|R_{12}|^2 + m_3|R_{13}|^2),$ where $R_{ij}$ represents the $(i,j)$ entry of the $R$ matrix.
We can now take $R_{13} = 0$, since there is no mixing in this component, and $m_1=0$ is experimentally allowed.
However, since we have $|R_{12}|^2 = 1/(1 + (M_1/M_2)\cot^2\theta_N) \sim 1/2$ in our setup with $M_1 / M_2 \ll 1$, large $M_1$ is not allowed because of the X-ray constraint $\Theta^2_{11} \simeq m_2/(2M_1)\lesssim 10^{-5}({\rm keV}/M_1)^5$, where $m_2\simeq \sqrt{\Delta m^2_{21}}$ in the NH case, and $m_2$ is replaced by $m_1\simeq\sqrt{|\Delta m^2_{32}|}$ in the IH case. 
One may naively expect that this decay of light RHN through a heavier RHN is suppressed by the hierarchically large mass ratio $M_1/M_2\ll 1$. If we did not enforce the simple seesaw mechanism to obtain the desirable light neutrino masses, this would be the case and the X-ray bound could be circumvented. We, however, in our model construction stick to the seesaw mechanism to account for the observed neutrino masses, which then inevitably increase $y_2$ if we choose a bigger value of $M_2$ to result in too big an X-ray decay rate. To keep the virtue of explaining the observed neutrino masses by the simple type-I seesaw mechanism and yet not to lose the attractive feature of simple RHN oscillation production, we now discuss a time-dependent RHN mixing to evade the X-ray constraint mentioned above.

Such a time-dependent RHN mixing can be achieved by utilizing the dynamics of a real scalar filed $\phi$.
Let us here consider the two flavor case for simplicity, but the extension to the three flavor system is straightforward.
In the two flavor case, we impose $Z_2$ symmetry under which $\nu_{R2}$ is even, while $\nu_{R1}$ and $\phi$ are odd.
\footnote{
Although our setup is similar to the idea discussed in Ref.~\cite{Berlin:2016bdv}, the DM production scenario is quite different, since our scenario does not rely on the oscillation between active and sterile neutrinos, and thus the temperature at which the production efficiently occurs takes rather a wide range, which can imprint an observable signature on the structure formation. 
}
Now the mass matrix ${\cal M}_N$ in Eq. (\ref{eq:lagrangian}) is given by
\begin{eqnarray}
	{\cal M}_N(\phi) =
	\left[
		\begin{array}{cc}
			M_1 & \kappa \phi\\
			\kappa \phi & M_2
		\end{array}
	\right]
\end{eqnarray}
in the interaction basis.
The dynamics of $\phi$ is governed by the equation of motion $\ddot{\phi} + 3H\dot{\phi} + V'(\phi) = 0$, where $V(\phi)$ is the potential that we take $V(\phi) \simeq (1/2)m_\phi^2\phi^2$.
For $m_\phi\ll 3H$, $\phi$ is almost constant, namely, $\phi \simeq \sqrt{2\rho_\phi}/m_\phi$ with $\rho_\phi$ the energy density of $\phi$, and when $H$ drops below $m_\phi$, $\phi$ starts to oscillate.
As we will see below, $m_\phi\ll 3H$ is always satisfied when the $N_1$ production rate is maximal, and thus we take $\phi$ as a constant in this regime.

The mixing angle between $\nu_{R1}$ and $\nu_{R2}$ is given by $\sin\theta_N \simeq \kappa \phi / M_2$ in the case that $M_1 \ll M_2$, and thus in the constant $\phi$ regime we obtain $\sin^22\theta_N \simeq 4\kappa^2 \rho_\phi/(m_\phi^2M_2^2)$, where the relevant $\theta_N$ is determined by $\rho_\phi(T_{\rm max})$ with $T_{\max}$ being the temperature at which the production rate becomes maximal, namely, $T_{\rm max} \sim T_c$ for $M_2\lesssim T_c$ and otherwise $T_{\rm max}\sim M_2$.
As mentioned above, $\phi$ is constant until it starts to oscillate, so we can take $\rho_\phi(T_{\rm max}) \simeq \rho_\phi(T_{\rm osc})$ with $T_{\rm osc}$ given by $m_\phi = 3H(T_{\rm osc})$.
Then, we obtain
\begin{eqnarray}
	\sin^22\theta_N &\simeq&
	0.3\times
	\left[
		\frac{r_g}{30}
	\right]^{1/4}
	\left[
		\frac{\kappa}{10^{-9}}
	\right]^2
	\left[
		\frac{m_\phi}{10^{-4}~{\rm eV}}
	\right]^{-1/2}\nonumber\\
	&&
	\times
	\left[
		\frac{M_2}{100~{\rm GeV}}
	\right]^{-2}
	\left[
		\frac{r}{10^{-4}}
	\right],\label{eq:phi-mixing}
\end{eqnarray}
with $r_g= g_*(T_{\rm osc})/g_*(T_0)$, and $r=\rho_\phi^0/\rho_{\rm DM}$ with $\rho_\phi^0$ and $\rho_{\rm DM}$ being the energy density of $\phi$ and dark matter at the present.
Here we have used $g_*(T_0)\simeq3.36$.

We also require that $\phi$ never thermalizes by taking a sufficiently small $\kappa$ not to affect the big bang nucleosynthesis, which results in $\kappa^2 \lesssim M_2/M_{\rm Pl}$.
In addition, $m_\phi$ should be smaller than $H(T_{\rm max})$ in order for $\phi$ at $T_{\rm max}$ to be constant, where $H(T_{\rm max})\simeq 10^{-5}$ eV for $M_2 < T_c$ and $H(T_{\rm max})\simeq 10^{-5}\times (M_2/T_c)^2$ for $M_2 > T_c$.

It is worth mentioning that the dynamics of $\phi$ may be tied to inflationary models.
In particular, the condition of $\rho_\phi^0\ll\rho_{\rm DM}$ implies that the initial amplitude of $\phi$ is bounded
\begin{eqnarray}
	\phi\lesssim 4\times10^{11}~{\rm GeV}\left(\frac{r_g}{30}\right)^{1/2}\left(\frac{r}{10^{-4}}\right)^{1/2}\left(\frac{10^{-4}~{\rm eV}}{m_\phi}\right)^{1/4}.
\end{eqnarray}
On the other hand, $\phi$ could be largely displaced from the origin during inflation and its oscillation at a later time possibly dominates the dark matter energy density, in an analogous manner to the Polonyi/moduli problem \cite{Coughlan:1983ci,Ellis:1986zt,Goncharov:1984qm}.
To suppress $\phi$ in our case, we may utilize a relatively strong coupling between $\phi$ and inflaton, which renders the adiabatic suppression of the amplitude of the coherent oscillations \cite{Linde:1996cx}.
Its actual dynamics, however, depends on the inflationary models and how $\phi$ couples to the inflaton, which we leave unspecified for the future work.

Finally let us comment on the $\theta_N$ at the present, which is relevant for the decay of $N_1$.
Below $T_{\rm osc}$, since $\rho_\phi$ drops as a matter energy density, we obtain
\begin{eqnarray}
	\frac{\sin^22\theta_N(T_0)}{\sin^22\theta_N(T_{\rm osc})} \simeq 1.2\times 10^{-46}
	\left[
		\frac{r_g}{30}
	\right]^{-1/4}
	\left[
		\frac{m_\phi}{10^{-4}~{\rm eV}}
	\right]^{-3/2},
\end{eqnarray}
and therefore a sufficiently small mixing to avoid the X-ray constraint can be achieved.

\section{Discussion/Conclusion}
Before concluding our discussions, let us briefly point out another
potentially interesting production mechanism: the production of $N_1$
from a heavier RHN decay. We can consider the decay of $N_2$ (and/or
$N_3$) which is thermally decoupled while it is relativistic
(otherwise $N_2$ number density would be too small due to the
Boltzmann suppression). $N_1$ abundance then can be estimated as
\begin{eqnarray}
\Omega_{N_1}h^2
&\simeq& 10^{-10}
\left[
\frac{\Theta^2_{11}}{10^{-12}}
\right]
\left[
\frac{M_1}{10~{\rm keV}}
\right]
\left[
\frac{g_*(T_0)}{g_*(T_{\rm FO})}
\right]
\end{eqnarray}
where we used the branching fraction of $N_2$ decay for the process
$N_2\to N_1+$(mesons, leptons), ${\rm Br}(N_2\to N_1)\simeq \Gamma(N_2
\to N_1)/\Gamma(N_2 \to SM) \simeq  M_2 \Theta^2_{11}\Theta^2_{22}/
M_2 \Theta^2_{22} \simeq \Theta^2_{11}$, and the ratio of $g_*$
accounts for the change in the effective degrees of freedom from the
$N_2$ freeze-out epoch to the present time. This production
contribution is hence subdominant compared with RHN oscillation
production in the parameter region of our interest.

Let us next mention the small-scale structure constraints applicable to our scenario. We here discuss the Lyman-$\alpha$ forest constraints which can give the lower limit on the DM mass from the DM free streaming scale $\lambda_{FS}\sim 1 ~\mbox{Mpc} ( {\rm keV}/{M_1}) ({\langle p/T \rangle}/{3.15})$ \cite{kev01}.
Too large a free streaming scale can be excluded due to the suppression of small-scale structure formation.
The average momentum of $N_1$ produced by the nonresonant oscillation of thermalized $N_2$ can be estimated as $\langle p_1\rangle \sim 2.8T$, analogous to the conventional (nonresonant) active-sterile oscillation scenario. Taking account of momentum redshifting by a factor $(g_*(T_{N_2\rightarrow N_1})/g_*(T\ll {\rm MeV}))^{-1/3}$ due to the change in the effective
degrees of freedom, Lyman-$\alpha$ data leads to the RHN DM mass bound $M_1 \gtrsim  10 $ keV for our scenario \cite{ir2017} (when $N_2\rightarrow N_1$ occurs most efficiently before the QCD phase transition which is the case for the parameter range discussed so far). Such a DM mass range can be realized in our scenario as explicitly demonstrated through the concrete examples in the last section while being compatible with both the right relic abundance and seesaw mechanism. 

Among the possible extensions of our DM scenarios, we plan to study the leptogenesis as well as the neutrino observables such as the neutrinoless double beta decay in our future work. For instance, even though we have focused on the DM production in this article, the neutrino Yukawa couplings in our model can be further constrained by seeking the production of desirable baryon asymmetry in the Universe. The realization of leptogenesis when $N_2$ and $N_3$ are heavy enough and/or are degenerate in their masses with sufficient $CP$ violations \cite{Fukugita:1986hr,Asaka:2005pn,Pilaftsis:2003gt} will be explored in our forthcoming paper. The $CP$ phases in the neutrino Yukawa couplings are of great importance not only for the leptogenesis but also for the DM production in our scenario, and the presented production mechanism for the RHN DM could uncover a new connection between DM and leptogenesis to bring considerable opportunities for subsequent studies. 

\begin{acknowledgments}
This work was supported by IBS under the project code IBS-R018-D1.
We thank A. Kamada, A. Merle and T. Asaka for useful discussions and, in particular, the anonymous referee for the constructive suggestions.
\end{acknowledgments}


\begin{thebibliography}{99}  

\bibitem{Fukuda:1998mi} 
  Y.~Fukuda {\it et al.} [Super-Kamiokande Collaboration],
  Phys.\ Rev.\ Lett.\  {\bf 81}, 1562 (1998)
  doi:10.1103/PhysRevLett.81.1562
  [hep-ex/9807003].

\bibitem{Ahmad:2002jz} 
  Q.~R.~Ahmad {\it et al.} [SNO Collaboration],
  Phys.\ Rev.\ Lett.\  {\bf 89}, 011301 (2002)
  doi:10.1103/PhysRevLett.89.011301
  [nucl-ex/0204008].
    
\bibitem{seesaw}
  T. ~Yanagida,
  in Proceedings of the Workshop on Unified Theory and Baryon Number of the Universe,
  edited by O. Sawada and A. Sugamoto [Conf. Proc. C7902131, p. 95–99 (1979)];
  M. Gell- Mann, P. Ramond and R. Slansky,
  in Supergravity,
  eds. P. van Niewwenhuizen and D. Freedman (North Holland, Amsterdam, 1979); 
  S.L. Glashow,
  in Quarks and Leptons, Carg\`{e}se 1979,
  eds. M. L\'{e}vy, et al., (Plenum 1980 New York), p. 707.
  See also 
  P.~Minkowski,
  Phys.\ Lett.\  {\bf B67}, 421 (1977).

\bibitem{Asaka:2005an} 
  T.~Asaka, S.~Blanchet and M.~Shaposhnikov,
  Phys.\ Lett.\ B {\bf 631}, 151 (2005)
  doi:10.1016/j.physletb.2005.09.070
  [hep-ph/0503065].

\bibitem{Fukugita:1986hr} 
  M.~Fukugita and T.~Yanagida,
  Phys.\ Lett.\ B {\bf 174}, 45 (1986).
  doi:10.1016/0370-2693(86)91126-3

\bibitem{Asaka:2005pn} 
  T.~Asaka and M.~Shaposhnikov,
  Phys.\ Lett.\ B {\bf 620}, 17 (2005)
  doi:10.1016/j.physletb.2005.06.020
  [hep-ph/0505013].
  
\bibitem{Canetti:2012kh} 
  L.~Canetti, M.~Drewes, T.~Frossard and M.~Shaposhnikov,
  Phys.\ Rev.\ D {\bf 87}, 093006 (2013)
  doi:10.1103/PhysRevD.87.093006
  [arXiv:1208.4607 [hep-ph]].

\bibitem{Ibe:2015nfa} 
  M.~Ibe and K.~Kaneta,
  Phys.\ Rev.\ D {\bf 92}, no. 3, 035019 (2015)
  doi:10.1103/PhysRevD.92.035019
  [arXiv:1504.04125 [hep-ph]].

\bibitem{kk2}
  H.~Murayama, H.~Suzuki, T.~Yanagida and J.~Yokoyama,
  Phys.\ Rev.\ D {\bf 50}, R2356 (1994)
  doi:10.1103/PhysRevD.50.R2356
  [hep-ph/9311326];
  J.~R.~Ellis, M.~Raidal and T.~Yanagida,
  Phys.\ Lett.\ B {\bf 581}, 9 (2004)
  doi:10.1016/j.physletb.2003.11.029
  [hep-ph/0303242];
  K.~Kadota and J.~Yokoyama,
  Phys.\ Rev.\ D {\bf 73}, 043507 (2006)
  doi:10.1103/PhysRevD.73.043507
  [hep-ph/0512221]; 
  
\bibitem{Dodelson:1993je} 
  S.~Dodelson and L.~M.~Widrow,
  Phys.\ Rev.\ Lett.\  {\bf 72}, 17 (1994)
  doi:10.1103/PhysRevLett.72.17
  [hep-ph/9303287].

\bibitem{Tremaine:1979we} 
  S.~Tremaine and J.~E.~Gunn,
  Phys.\ Rev.\ Lett.\  {\bf 42}, 407 (1979).
  doi:10.1103/PhysRevLett.42.407

\bibitem{Boyarsky:2005us} 
  A.~Boyarsky, A.~Neronov, O.~Ruchayskiy and M.~Shaposhnikov,
  Mon.\ Not.\ Roy.\ Astron.\ Soc.\  {\bf 370}, 213 (2006)
  doi:10.1111/j.1365-2966.2006.10458.x
  [astro-ph/0512509].
  
\bibitem{Horiuchi:2013noa} 
  S.~Horiuchi, P.~J.~Humphrey, J.~Onorbe, K.~N.~Abazajian, M.~Kaplinghat and S.~Garrison-Kimmel,
  Phys.\ Rev.\ D {\bf 89}, no. 2, 025017 (2014)
  doi:10.1103/PhysRevD.89.025017
  [arXiv:1311.0282 [astro-ph.CO]];
  K.~Perez, K.~C.~Y.~Ng, J.~F.~Beacom, C.~Hersh, S.~Horiuchi and R.~Krivonos,
  Phys.\ Rev.\ D {\bf 95}, no. 12, 123002 (2017)
  doi:10.1103/PhysRevD.95.123002
  [arXiv:1609.00667 [astro-ph.HE]].
  J.~F.~Cherry and S.~Horiuchi,
  Phys.\ Rev.\ D {\bf 95}, no. 8, 083015 (2017)
  doi:10.1103/PhysRevD.95.083015
  [arXiv:1701.07874 [hep-ph]].

\bibitem{ir2017} 
  V.~Iršič {\it et al.},
  Phys.\ Rev.\ D {\bf 96}, no. 2, 023522 (2017)
  doi:10.1103/PhysRevD.96.023522
  [arXiv:1702.01764 [astro-ph.CO]].

\bibitem{Shi:1998km} 
  X.~D.~Shi and G.~M.~Fuller,
  Phys.\ Rev.\ Lett.\  {\bf 82}, 2832 (1999)
  doi:10.1103/PhysRevLett.82.2832
  [astro-ph/9810076].

\bibitem{Asaka:2006ek}
  T.~Asaka, M.~Shaposhnikov and A.~Kusenko,
  Phys.\ Lett.\ B {\bf 638}, 401 (2006)
  doi:10.1016/j.physletb.2006.05.067
  [hep-ph/0602150];
  M.~Shaposhnikov and I.~Tkachev,
  Phys.\ Lett.\ B {\bf 639}, 414 (2006)
  doi:10.1016/j.physletb.2006.06.063
  [hep-ph/0604236];
  A.~Kusenko,
  Phys.\ Rev.\ Lett.\  {\bf 97}, 241301 (2006)
  doi:10.1103/PhysRevLett.97.241301
  [hep-ph/0609081];
  K.~Petraki and A.~Kusenko,
  Phys.\ Rev.\ D {\bf 77}, 065014 (2008)
  doi:10.1103/PhysRevD.77.065014
  [arXiv:0711.4646 [hep-ph]].
  H.~Matsui and M.~Nojiri,
  Phys.\ Rev.\ D {\bf 92}, no. 2, 025045 (2015)
  doi:10.1103/PhysRevD.92.025045
  [arXiv:1503.01293 [hep-ph]];
  A.~Merle, V.~Niro and D.~Schmidt,
  JCAP {\bf 1403}, 028 (2014)
  doi:10.1088/1475-7516/2014/03/028
  [arXiv:1306.3996 [hep-ph]];
  Z.~Kang,
  Eur.\ Phys.\ J.\ C {\bf 75}, no. 10, 471 (2015)
  doi:10.1140/epjc/s10052-015-3702-4
  [arXiv:1411.2773 [hep-ph]];
  S.~B.~Roland, B.~Shakya and J.~D.~Wells,
  Phys.\ Rev.\ D {\bf 92}, no. 11, 113009 (2015)
  doi:10.1103/PhysRevD.92.113009
  [arXiv:1412.4791 [hep-ph]];
  A.~Merle and M.~Totzauer,
  JCAP {\bf 1506}, 011 (2015)
  doi:10.1088/1475-7516/2015/06/011
  [arXiv:1502.01011 [hep-ph]];
  Z.~Kang,
  Phys.\ Lett.\ B {\bf 751}, 201 (2015)
  doi:10.1016/j.physletb.2015.10.031
  [arXiv:1505.06554 [hep-ph]];
  A.~Adulpravitchai and M.~A.~Schmidt,
  JHEP {\bf 1512}, 023 (2015)
  doi:10.1007/JHEP12(2015)023
  [arXiv:1507.05694 [hep-ph]];
  M.~Drewes and J.~U.~Kang,
  JHEP {\bf 1605}, 051 (2016)
  doi:10.1007/JHEP05(2016)051
  [arXiv:1510.05646 [hep-ph]].

\bibitem{Bezrukov:2009th} 
  F.~Bezrukov, H.~Hettmansperger and M.~Lindner,
  Phys.\ Rev.\ D {\bf 81}, 085032 (2010)
  doi:10.1103/PhysRevD.81.085032
  [arXiv:0912.4415 [hep-ph]];
  M.~Nemevsek, G.~Senjanovic and Y.~Zhang,
  JCAP {\bf 1207}, 006 (2012)
  doi:10.1088/1475-7516/2012/07/006
  [arXiv:1205.0844 [hep-ph]];
  K.~Kaneta, Z.~Kang and H.~S.~Lee,
  JHEP {\bf 1702}, 031 (2017)
  doi:10.1007/JHEP02(2017)031
  [arXiv:1606.09317 [hep-ph]].

\bibitem{kenji} 
  K.~Kadota,
  Phys.\ Rev.\ D {\bf 77}, 063509 (2008)
  doi:10.1103/PhysRevD.77.063509
  [arXiv:0711.1570 [hep-ph]]; 
  A.~V.~Patwardhan, G.~M.~Fuller, C.~T.~Kishimoto and A.~Kusenko,
  Phys.\ Rev.\ D {\bf 92}, no. 10, 103509 (2015)
  doi:10.1103/PhysRevD.92.103509
  [arXiv:1507.01977 [astro-ph.CO]]; 
  A.~Merle, A.~Schneider and M.~Totzauer,
  JCAP {\bf 1604}, no. 04, 003 (2016)
  doi:10.1088/1475-7516/2016/04/003
  [arXiv:1512.05369 [hep-ph]].

\bibitem{Anisimov:2008gg} 
  A.~Anisimov and P.~Di Bari,
  Phys.\ Rev.\ D {\bf 80}, 073017 (2009)
  doi:10.1103/PhysRevD.80.073017
  [arXiv:0812.5085 [hep-ph]];
  P.~Di Bari, P.~O.~Ludl and S.~Palomares-Ruiz,
  JCAP {\bf 1611}, no. 11, 044 (2016)
  doi:10.1088/1475-7516/2016/11/044
  [arXiv:1606.06238 [hep-ph]].

\bibitem{Adhikari:2016bei}
  M.~Drewes {\it et al.},
  JCAP {\bf 1701}, no. 01, 025 (2017)
  doi:10.1088/1475-7516/2017/01/025
  [arXiv:1602.04816 [hep-ph]].

\bibitem{Casas:2001sr} 
  J.~A.~Casas and A.~Ibarra,
  Nucl.\ Phys.\ B {\bf 618}, 171 (2001)
  doi:10.1016/S0550-3213(01)00475-8
  [hep-ph/0103065]:
  A.~Broncano, M.~B.~Gavela and E.~E.~Jenkins,
  Phys.\ Lett.\ B {\bf 552}, 177 (2003)
  Erratum: [Phys.\ Lett.\ B {\bf 636}, 332 (2006)]
  doi:10.1016/j.physletb.2006.04.003, 10.1016/S0370-2693(02)03130-1
  [hep-ph/0210271];
  J.~A.~Casas, A.~Ibarra and F.~Jimenez-Alburquerque,
  JHEP {\bf 0704}, 064 (2007)
  doi:10.1088/1126-6708/2007/04/064
  [hep-ph/0612289];
  M.~Blennow and E.~Fernandez-Martinez,
  Phys.\ Lett.\ B {\bf 704}, 223 (2011)
  doi:10.1016/j.physletb.2011.09.028
  [arXiv:1107.3992 [hep-ph]];
  J.~Heeck,
  Phys.\ Rev.\ D {\bf 86}, 093023 (2012)
  doi:10.1103/PhysRevD.86.093023
  [arXiv:1207.5521 [hep-ph]].

\bibitem{Gonzalez-Garcia:2015qrr} 
  M.~C.~Gonzalez-Garcia, M.~Maltoni and T.~Schwetz,
  Nucl.\ Phys.\ B {\bf 908}, 199 (2016)
  doi:10.1016/j.nuclphysb.2016.02.033
  [arXiv:1512.06856 [hep-ph]].

 \bibitem{Dolgov:2000ew} 
  A.~D.~Dolgov and S.~H.~Hansen,
  Astropart.\ Phys.\  {\bf 16}, 339 (2002)
  doi:10.1016/S0927-6505(01)00115-3
  [hep-ph/0009083].

\bibitem{Pilaftsis:2003gt} 
  A.~Pilaftsis and T.~E.~J.~Underwood,
  Nucl.\ Phys.\ B {\bf 692}, 303 (2004)
  doi:10.1016/j.nuclphysb.2004.05.029
  [hep-ph/0309342].

\bibitem{Besak:2012qm} 
  D.~Besak and D.~Bodeker,
  JCAP {\bf 1203}, 029 (2012)
  doi:10.1088/1475-7516/2012/03/029
  [arXiv:1202.1288 [hep-ph]].

\bibitem{DOnofrio:2014rug} 
  M.~D'Onofrio, K.~Rummukainen and A.~Tranberg,
  Phys.\ Rev.\ Lett.\  {\bf 113}, no. 14, 141602 (2014)
  doi:10.1103/PhysRevLett.113.141602
  [arXiv:1404.3565 [hep-ph]].

\bibitem{DOnofrio:2015gop} 
  M.~D'Onofrio and K.~Rummukainen,
  Phys.\ Rev.\ D {\bf 93}, no. 2, 025003 (2016)
  doi:10.1103/PhysRevD.93.025003
  [arXiv:1508.07161 [hep-ph]].
  
\bibitem{Akhmedov:2012uu} 
  E.~Akhmedov, D.~Hernandez and A.~Smirnov,
  JHEP {\bf 1204}, 052 (2012)
  doi:10.1007/JHEP04(2012)052
  [arXiv:1201.4128 [hep-ph]].

\bibitem{Ade:2015xua} 
  P.~A.~R.~Ade {\it et al.} [Planck Collaboration],
  Astron.\ Astrophys.\  {\bf 594}, A13 (2016)
  doi:10.1051/0004-6361/201525830
  [arXiv:1502.01589 [astro-ph.CO]].

\bibitem{Berlin:2016bdv} 
  A.~Berlin and D.~Hooper,
  Phys.\ Rev.\ D {\bf 95}, no. 7, 075017 (2017)
  doi:10.1103/PhysRevD.95.075017
  [arXiv:1610.03849 [hep-ph]].

\bibitem{Coughlan:1983ci} 
  G.~D.~Coughlan, W.~Fischler, E.~W.~Kolb, S.~Raby and G.~G.~Ross,
  Phys.\ Lett.\  {\bf 131B}, 59 (1983).
  doi:10.1016/0370-2693(83)91091-2

\bibitem{Ellis:1986zt} 
  J.~R.~Ellis, D.~V.~Nanopoulos and M.~Quiros,
  Phys.\ Lett.\ B {\bf 174}, 176 (1986).
  doi:10.1016/0370-2693(86)90736-7

\bibitem{Goncharov:1984qm} 
  A.~S.~Goncharov, A.~D.~Linde and M.~I.~Vysotsky,
  Phys.\ Lett.\  {\bf 147B}, 279 (1984).
  doi:10.1016/0370-2693(84)90116-3

\bibitem{Linde:1996cx} 
  A.~D.~Linde,
  Phys.\ Rev.\ D {\bf 53}, R4129 (1996)
  doi:10.1103/PhysRevD.53.R4129
  [hep-th/9601083].


\bibitem{kev01} 
  K.~Abazajian, G.~M.~Fuller and M.~Patel,
  Phys.\ Rev.\ D {\bf 64}, 023501 (2001)
  doi:10.1103/PhysRevD.64.023501
  [astro-ph/0101524].
\end{thebibliography}
\end{document}